\documentclass[
reprint,
superscriptaddress,
nonbibnotes,
longbibliography,
floatfix,
amsmath,
amssymb,
aps,
prb
]{revtex4-2}
\usepackage[english]{babel}
\usepackage{graphicx}
\usepackage{soul}
\usepackage{siunitx}
\usepackage{lineno}
\usepackage[version=4]{mhchem}
\usepackage[breaklinks=true,colorlinks=true,linkcolor=blue,urlcolor=blue,citecolor=blue]{hyperref}
\usepackage[utf8]{inputenc}
\renewcommand{\selectlanguage}[1]{}  %resolves problem in babel with "language=en" or similar.

\begin{document}

\title{Creating functional nanostructures for vortex manipulation in superconducting {Bi$_2$Sr$_2$CaCu$_2$O$_{8+x}$} thin films via focused He ion beam irradiation}

%\date{\today}

\author{Bernd Aichner} \email{bernd.aichner@univie.ac.at}
%\thanks{\equalcontribution}
\affiliation{Faculty of Physics, University of Vienna, 1090 Vienna, Austria}

\author{Christoph Schmid}
%\thanks{\equalcontribution}
\affiliation{Physikalisches Institut, Center for Quantum Science (CQ) and LISA$^+$, Universität Tübingen, 72076 Tübingen, Germany}

\author{Sandra Keppert}
%\thanks{\equalcontribution}
\affiliation{Institute of Applied Physics, Johannes Kepler University Linz, 4040 Linz, Austria}

\author{Reinhold Kleiner}
\affiliation{Physikalisches Institut, Center for Quantum Science (CQ) and LISA$^+$, Universität Tübingen, 72076 Tübingen, Germany}

\author{Edward Goldobin}
\affiliation{Physikalisches Institut, Center for Quantum Science (CQ) and LISA$^+$, Universität Tübingen, 72076 Tübingen, Germany}

\author{Johannes D. Pedarnig}
\affiliation{Institute of Applied Physics, Johannes Kepler University Linz, 4040 Linz, Austria}

\author{Dieter Koelle}
\affiliation{Physikalisches Institut, Center for Quantum Science (CQ) and LISA$^+$, Universität Tübingen, 72076 Tübingen, Germany}

\author{Wolfgang Lang} \email{wolfgang.lang@univie.ac.at}
\affiliation{Faculty of Physics, University of Vienna, 1090 Vienna, Austria}

\begin{abstract}
Bi$_2$Sr$_2$CaCu$_2$O$_{8+x}$ (Bi-2212) is a highly anisotropic high-temperature superconductor exhibiting unique vortex dynamics due to its layered structure. Precise nanoscale patterning of ultrathin Bi-2212 films remains challenging, limiting controlled studies of vortex pinning. Here, we employ \qty{30}{\keV} focused helium ion beam (He-FIB) irradiation to fabricate periodic arrays of columnar defects with \qty{100}{\nm} spacings in optimally-doped Bi-2212 thin films, thereby creating artificial vortex pinning centers. Magnetotransport measurements reveal pronounced vortex matching effects, evidenced by minima in magnetoresistance and peaks in critical current at commensurate magnetic fields. The angular dependence of magnetoresistance and critical current scale with an effective magnetic field, indicating strong confinement of pancake vortices along irradiation-induced tracks. These findings demonstrate the efficacy of He-FIB nanopatterning in modulating the superconducting properties in Bi-2212 thin films, providing a platform for fundamental studies of vortex matter and advancing prospects for device applications based on this highly anisotropic superconductor.
\end{abstract}

%\keywords{Bi-2212 thin films, helium ion beam nanopatterning, vortex pinning, magnetoresistance, critical current density}

\maketitle

\section{Introduction}

The high-temperature superconductor {Bi$_2$Sr$_2$CaCu$_2$O$_{8+x}$} (Bi-2212) has long been a subject of intense research due to its high critical temperature $T_\mathrm{c} \sim \qty{90}{\K}$ and its highly anisotropic, layered crystal structure \cite{MAED96M}. This anisotropy manifests itself in the large ratio of out-of-plane to in-plane normal-state resistivities, $\rho_c(\qty{100}{\K})/\rho_{ab}(\qty{100}{\K}) \sim \num{2.2e4}$ \cite{HEIN99}, and in the superconducting anisotropy $\xi_{ab}/\xi_c = \lambda_c/\lambda_{ab} \sim 180$ \cite{AICH25}. Here, $\xi_{ab}$ and $\xi_{c}$ denote the Ginzburg-Landau coherence lengths at temperature $T=\qty{0}{\K}$ parallel and perpendicular to the CuO$_2$ layers of Bi-2212, respectively, and $\lambda_{ab}$ and $\lambda_{c}$ are the zero-temperature London penetration depths for magnetic fields oriented parallel to the crystallographic $c$ axis and along the $ab$ planes, respectively.

These peculiar properties lead to unique vortex physics where magnetic flux lines are better described as stacks of two-dimensional (2D) `pancake' vortices residing in the CuO$_2$ planes \cite{SAFA92b}, weakly coupled along the $c$ axis by Josephson interaction \cite{BLAT94R, KLEI92}. While much foundational work on Bi-2212 was conducted on bulk crystals \cite{KLEI92,HEIN99,WANG01a} and films with thicknesses of at least \qty{100}{nm} decades ago \cite{RAFF91a,RI94,KONS99,DURR00,ROSS01}, there has been a recent resurgence of interest in ultrathin Bi-2212, driven by the observation of strict 2D high-$T_\mathrm{c}$ superconductivity in monolayers \cite{YU19a} and its potential to assemble twisted van-der-Waals stacks of individual layers, unleashing a wealth of novel physical phenomena \cite{ZHAO23}. Moreover, a pronounced Hall-effect anomaly in few-layer Bi-2212 \cite{ZHAO19} is currently being debated \cite{GUO22,GUO26,PUIC04} and might be addressed by introducing artificially engineered vortex-pinning sites \cite{ZECH18a}.

However, the fabrication, manipulation, and, above all, the patterning of ultrathin Bi-2212 whiskers present a formidable challenge. The material is prone to rapid degradation in ambient atmosphere \cite{HUAN22} and requires sophisticated preparation techniques that preclude the fabrication of larger device structures \cite{SHOK24}. In contrast, very thin films comprising only a few Bi-2212 layers exhibit stable superconducting properties in a dry atmosphere for many months  \cite{AICH25}. Such thin films are crucial for both fundamental studies and applications such as superconducting single-photon detectors (SSPDs), where precise control over material properties and geometry is essential.

Focused He ion beam (He-FIB) irradiation has emerged as a powerful tool \cite{WARD06,HLAW16M} for creating structures in various materials \cite{HOFL23R}. For cuprate superconductors, most studies have focused on thin YBa$_{2}$Cu$_{3}$O$_{7-\delta}$ (YBCO) films, using He-FIB irradiation to create single Josephson junctions (JJs) \cite{CYBA15,MULL19,KARR24,Schm2025a}, Josephson diodes \cite{SCHM25}, JJ arrays \cite{LEFE19,PROP25}, and superconducting quantum interference devices (SQUIDs) \cite{CHO15,MULL19,Schm2025a}. He-FIB has also been applied to other superconductors, such as \ce{MgB2} \cite{KASA18}, NbN \cite{LI23}, NbTiN \cite{RUHT25}, and iron-based superconductors \cite{CHEN24}.  In the case of Bi-2212, applications of He-FIB have been mostly limited to flakes, where JJs \cite{WANG21} and single-photon detection \cite{CHAR23,MERI23} have been reported.

\begin{figure*}[!th]
    \centering
    \includegraphics[width=0.25\linewidth]{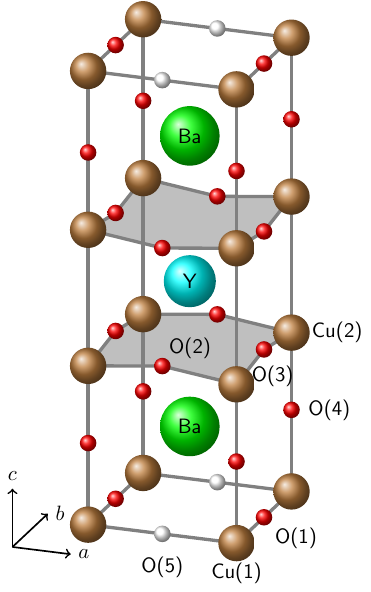}
    \includegraphics[width=0.25\linewidth]{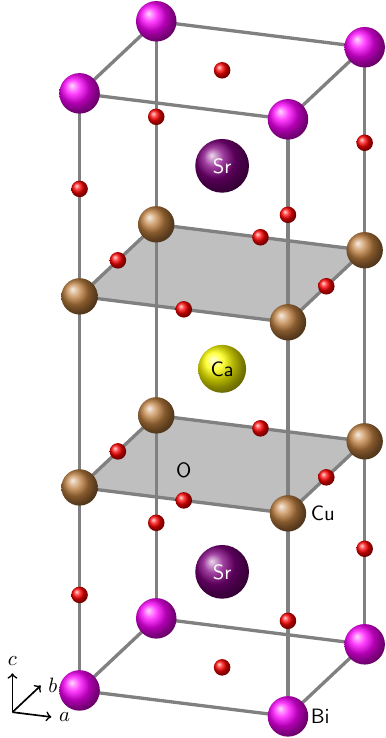}
    \caption{The crystal structures of YBCO (left) and Bi-2212 (right), shown in their stoichiometric representations. Atoms are depicted by individual colors and the copper-oxide layers are shaded in gray. White spheres indicate empty oxygen positions. Crystallographic axes are plotted to scale for comparison. For Bi-2212 a half unit cell is shown.}
    \label{fig:structure}
\end{figure*}

The effectiveness of He-FIB irradiation on larger areas can be evaluated by artificial vortex pinning landscapes that can be created via local suppression of $T_\mathrm{c}$. In YBCO films with pinning arrays induced by He ion irradiation, pronounced vortex pinning effects have been demonstrated for magnetic fields up to \qty{6}{T} \cite{KARR24a} and in complex geometries \cite{AICH19}, exhibiting robust temporal stability over time scales of several years \cite{KEPP24}. 

This work aims to explore the feasibility of He-FIB for functional nanopatterning of very thin $c$-axis oriented Bi-2212 films. As a model system, we investigate a periodic array of pinning sites through detailed magnetotransport measurements, including the angular dependences of magnetoresistance and  critical current density. By elucidating the specific characteristics of vortex pinning in these engineered Bi-2212 films and comparing them with established results for YBCO, we address the knowledge gap regarding how extreme anisotropy and pancake vortex dynamics influence pinning efficiency.

%---------------------------------
\section{Results and discussion}

\subsection{He-FIB Patterning}

The structural similarity between Bi-2212 and YBCO is illustrated in Fig.~\ref{fig:structure}. Both compounds have two \ce{CuO2} layers per unit cell (YBCO) or per half unit cell (Bi-2212), respectively. The unit cell of Bi-2212 consists of two subcells stacked along the $c$ axis and shifted diagonally by $(a/2, b/2)$, each containing one formula unit \cite{MAED96M}. The key difference compared to YBCO is the absence of Cu(1)O(1) chains in Bi-2212, thus forming a stack of \ce{BiO}, \ce{SrO}, \ce{CuO2}, Ca, \ce{CuO2}, \ce{SrO}, and \ce{BiO} layers. In YBCO, the oxygen O(1) atoms of the chains are weakly bound by an estimated threshold displacement energy of $\sim \qty{2.8}{\eV}$ \cite{TOLP96} and are consequently particularly sensitive to relocation by electron \cite{TOLP96a} or ion irradiation \cite{VALL89,LANG09,PEDA10}, optical illumination \cite{MARK96},  electromigration \cite{MOEC93,MARI20}, and preparation processes \cite{WANG95b}. Reordering the chain oxygen atoms to either the initially empty O(5) sites or to interstitial positions significantly weakens or destroys superconductivity locally \cite{TOLP96}, though leaving the mobile carrier concentration essentially unaffected \cite{AICH26}. Unfortunately, the binding energies of oxygen atoms in Bi-2212 are less well characterized. However, one might speculate that they are bound similarly to the O(2) atoms in YBCO ($\sim \qty{8.4}{\eV}$ \cite{TOLP96}) since they have an analogous coordination with neighboring atoms. The Kinchin--Pease model \cite{KINC55R} predicts that the number of displaced atoms is inversely proportional to their binding energies and directly proportional to the fluence of the incident ions. On this basis, one may hypothesize that, in Bi-2212, an ion fluence approximately three times higher than that required for YBCO is necessary to achieve comparable oxygen defect densities.

A critical caveat is that the chemical composition of real Bi-2212 is substantially more complex than implied by the idealized stoichiometry {Bi$_2$Sr$_2$CaCu$_2$O$_{8+x}$}. Hole doping is introduced predominantly via interstitial (excess) oxygen, parameterized by $x$, which resides in or in the vicinity of the \ce{SrO} and \ce{BiO} layers. These oxygen species are characterized by a relatively low displacement threshold energy of approximately \qty{0.15}{eV} \cite{TORS18}. In addition, Bi-2212 thin films often display appreciable cation nonstoichiometry, most notably \ce{Bi^3+} substitution on \ce{Sr^2+} sites \cite{EISA04}, which can also contribute to hole doping. Bi-2212 further exhibits an incommensurate modulation of the $c$-axis lattice parameter \cite{MAED96M}. Regardless of these complexities, it is reasonable to assume that oxygen disorder within the \ce{CuO2} planes constitutes a principal mechanism for the suppression of superconductivity.

\begin{figure*}
    \centering
    \includegraphics[width=0.9\linewidth]{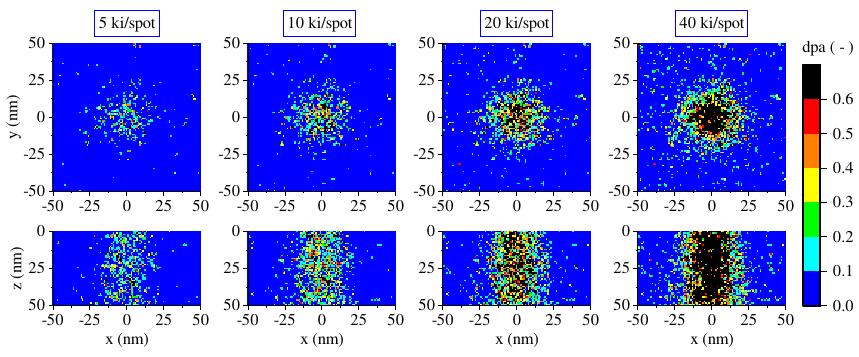}
    \caption{SRIM/TRIM simulations of defect channels created in Bi-2212 by He-FIB irradiation, with increasing fluence from left to right. The displacements per atom (dpa) in the film are shown color-coded in top views (cut at $z = \qty{25}{\nano\metre}$, upper panels) and side views (cut at $y = \qty{0}{\nano\metre}$, lower panels).}
    \label{fig:SRIM}
\end{figure*}

To test our hypothesis, we first performed SRIM/TRIM simulations \cite{ZIEG10} using a \qty{30}{\keV} \ce{He^+} ion beam with a Gaussian fluence profile of \qty{21}{\nm} full-width at half maximum (FWHM), incident on a \qty{50}{\nano\metre} thick Bi-2212 film while neglecting the substrate material. The results for various total numbers of orthogonally impacting ions are shown in Fig.~\ref{fig:SRIM}.   

For YBCO, the analysis of the relation between the displacements per atom (dpa) from SRIM/TRIM simulations and $T_\mathrm{c}$, determined experimentally from \qty{30}{\keV} \ce{He^+}-ion broad-field irradiation, revealed that superconductivity is fully suppressed at $\mathrm{dpa} \gtrsim 0.3$ \cite{AICH26}. For Bi-2212, equivalent data are not available. However, the structural equivalence of two copper-oxide layers that host superconductivity might suggest a similar value. This notion is supported by the finding in YBCO that the reduction of the critical temperature $\Delta T_\mathrm{c}$ is proportional to the average number of defects produced for any particle, energy, and fluence \cite{SUMM89}.  In this scenario, irradiation with \qty{5}{ki/spot} (5000 ions per spot) is expected to cause only minor local $T_\mathrm{c}$ depression, while \qty{40}{ki/spot} might provide a robust cylindrical pinning site with about \qty{30}{\nm} diameter. The evolution of such defect channels with increasing ion fluence is illustrated in Fig.~\ref{fig:SRIM}. 

\subsection{Magnetoresistance and Vortex Matching}

\begin{figure*}
    \centering
    \includegraphics[width=\linewidth]{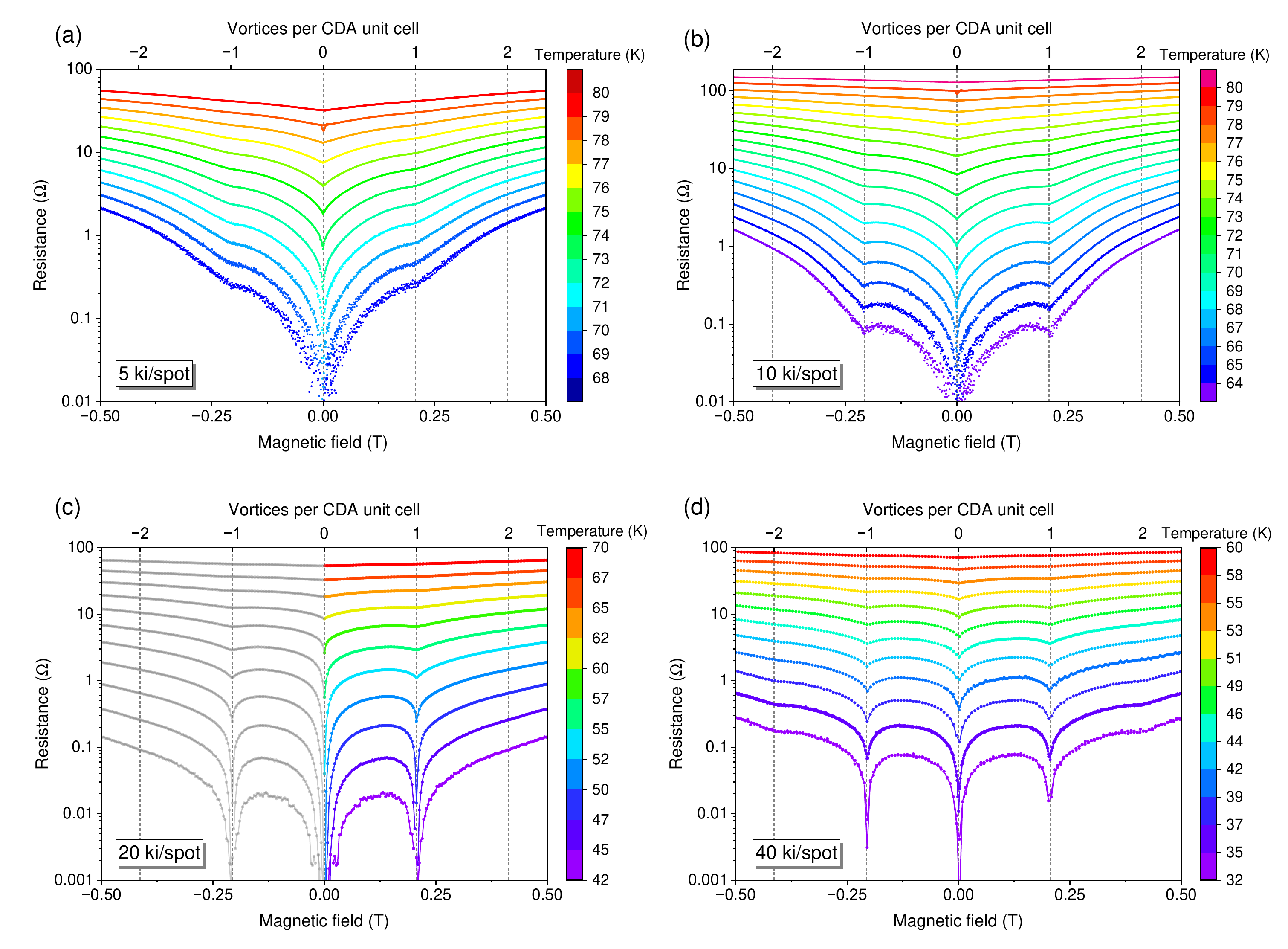}
    \caption{Magnetoresistance $R(B)$ of four different Bi-2212 microbridges containing a square lattice ($a=\qty{100}{\nm}$) of columnar defects, induced by He-FIB-irradiation with different fluences. The magnetic field $B$ was applied parallel to the crystallographic $c$ axis at various temperatures. The top scales indicate the number of vortices per unit cell of the CDA. In (c) data are only available for $B\ge\qty{0}{\T}$; corresponding $R$ values for negative $B$ are displayed in gray as a mirrored representation solely to enhance visual clarity.}
    \label{fig:RB}
\end{figure*}

In order to provide experimental support for the suggested suppression of $T_\mathrm{c}$ at high dpa also in Bi-2212, we irradiated several test bridges in a helium ion microscope to create a square array of columnar defects (CDs) with lattice constant $a = \qty{100}{\nm}$. Hence, the unit cell of the columnar-defect array (CDA) spans an area of $a^2$. For details on sample fabrication and He ion irradiation, see the methods section. These CDs act as effective pinning centers for Abrikosov vortices and can give rise to commensurability phenomena between the CDA and the vortex lattice, provided that inter-vortex interactions are substantial, i.e., $\lambda_{ab} \gtrsim a$. For comparable Bi-2212 thin films, we previously determined $\lambda_{ab} \sim \qty{200}{\nm}$ \cite{AICH25}. Under these conditions, the vortex-matching regime can immobilize vortices with respect to the applied Lorentz force, leading to dips in the resistance $R(B)$ and peaks in the critical current $I_c(B)$ for an applied magnetic field $B$ perpendicular to the film plane at a well-defined first matching field $B_\mathrm{m} = \Phi_0/a^2$, where $\Phi_0$ is the magnetic flux quantum. Note that matching effects can also occur at fields $k B_\mathrm{m}$, where $k$ is a rational number, but are observed more commonly at integer $k$. 

All bridges exhibited superconducting transitions to zero resistance after irradiation. However, due to minor deflections of impacting \ce{He^+} ions from their trajectories and recoiling atoms from the target material, some defects are expected to emerge outside of the CDs, as can be noticed in Fig.~\ref{fig:SRIM}. This leads to a reduction of the overall $T_\mathrm{c}$, defined by the inflection point of the $R(T)$ curve at the superconducting transition. In our samples,the reduction of $T_\mathrm{c}$ upon irradiation was $\Delta T_\mathrm{c} \sim \qty{-0.9}{\K}$ for \qty{5}{ki/spot}, \qty{-4.4}{\K} for \qty{10}{ki/spot}, \qty{-5.5}{\K} for \qty{20}{ki/spot}, and \qty{-14.5}{\K} for \qty{40}{ki/spot}. 

Fig.~\ref{fig:RB} shows representative magnetoresistance $R(B)$ curves ($I$ and $V$ are along the $a$ axis; $B$ along the $c$ axis) at various temperatures for four optimally-doped, \qtyrange{42}{49}{\nm} thick Bi-2212 bridges that were He-FIB-irradiated with different fluences. While for the sample irradiated with \qty{5}{ki/spot} $R$ increases monotonically with $B$, after \qty{10}{ki/spot} irradiation, weak minima in $R(B)$ are observed. By contrast, in YBCO, \qty{10}{ki/spot} irradiation already produces pronounced matching effects, and $R(B)$ minima in 30-nm-thin YBCO films are already visible for \qty{2}{ki/spot} \cite{KARR24a}. The pinning strength of the CD array becomes substantially stronger with more intense irradiation, as demonstrated by pronounced magnetoresistance minima at $B_\mathrm{m} = \qty{0.207}{T}$, corresponding to the emergence of vortex matching effects in the bridges with \num{20} and \qty{40}{ki/spot} irradiation. In the latter case, a tiny depression at $2 B_\mathrm{m} = \qty{0.414}{\T}$ can be observed, indicating the trapping of two flux quanta per unit cell of the CDA.  These matching features become less pronounced at higher temperatures as thermal depinning effects become more dominant.

The emergence of matching effects in Bi-2212 appears to require higher ion numbers per spot than were required for YBCO, where typically \qty{10}{ki/spot} led to pronounced resistance dips \cite{KARR24a}. We attribute this to the absence of CuO chains in Bi-2212, which are known to be particularly sensitive to oxygen displacement and disordering by ion impact in YBCO \cite{GRAY22}. The primary mechanism for defect creation by \ce{He^+} ions in Bi-2212 is yet unknown, but it likely involves displacement of various atoms within the \ce{BiO}, \ce{SrO}, and \ce{CuO2} layers. The higher fluence requirement could be attributed to higher displacement energies for relevant atoms in Bi-2212, a more efficient dynamic annealing of defects during irradiation, or a different nature of the defects that contribute to pinning.

\subsection{Angular Dependence of Magnetoresistance}

\begin{figure*}
    \centering
    \includegraphics[width=0.9\linewidth]{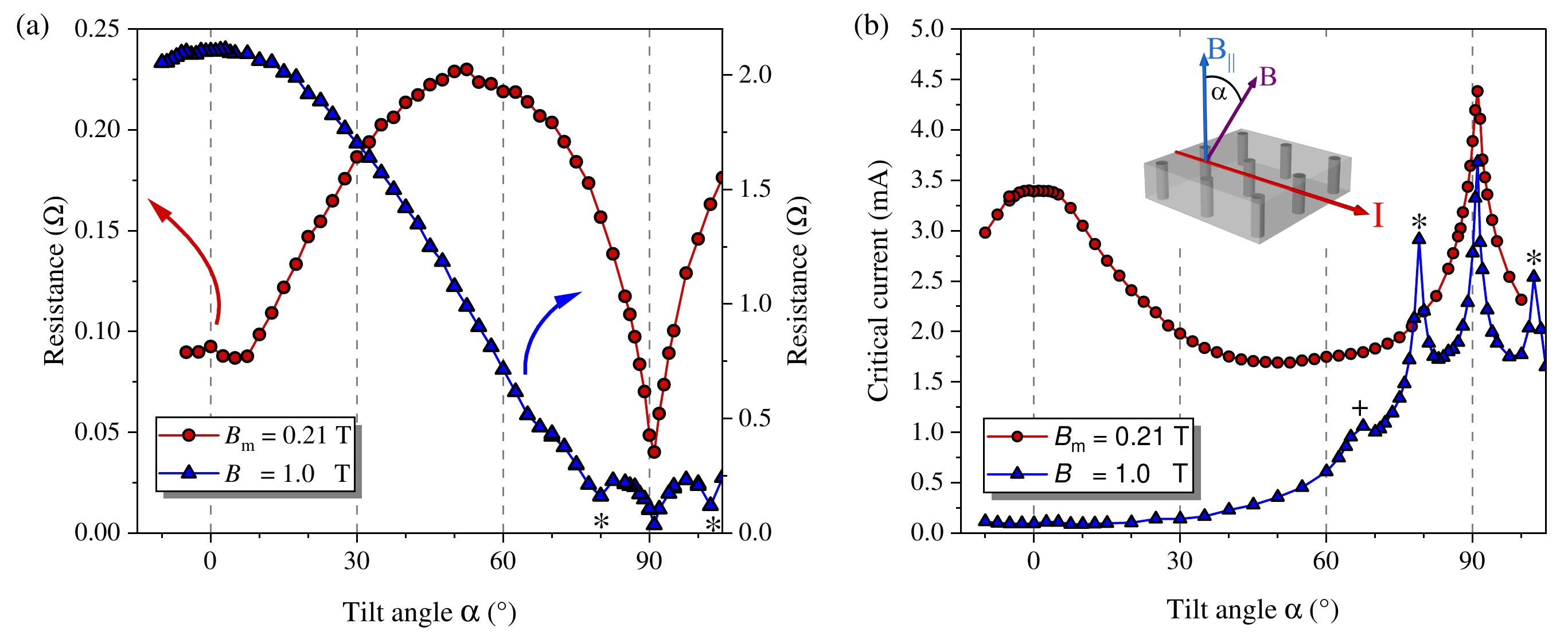}
    \caption{Angular dependence of transport properties on magnetic field direction in a Bi-2212 bridge with the pinning sites irradiated with \qty{40}{ki/spot} and measured at fixed $T$ for two values of applied $B=B_\mathrm{m}$ (dots) and $B=\qty{1.0}{\T}$ (triangles). Symbols (*, +) mark commensurability effects (see text). (a) $R(\alpha)$ at $T = \qty{33}{\K}$. (b) $I_c(\alpha)$ at $T = \qty{20}{\K}$. The inset shows the geometry of the measurement.}
    \label{fig:tilt}
\end{figure*}

The highly anisotropic nature of Bi-2212 significantly influences its response to tilted magnetic fields \cite{CURR18}. Fig.~\ref{fig:tilt} illustrates the resistance $R(\alpha)$ [Fig.~\ref{fig:tilt}(a)] and critical current $I_c(\alpha)$ [Fig.~\ref{fig:tilt}(b)] of the Bi-2212 bridge, nanopatterned with \qty{40}{ki/spot}, as a function of the tilt angle $\alpha$ between the applied magnetic field $B$ and the $c$ axis. The magnetic field is rotated in a plane orthogonal to the current [see the inset in Fig.~\ref{fig:tilt}(b)]. The behavior of Bi-2212 at two different field strengths $B=B_\mathrm{m}$ and a substantially higher $B=\qty{1.0}{\T}$ is compared. 

Figure \ref{fig:tilt}(a) shows $R(\alpha)$ at $T = \qty{33}{\K}$. At sufficiently high magnetic fields, where the vortex lattice becomes dense and any commensurability (matching) effects, if present, are expected to be negligible ($B=\qty{1.0}{\T}$; blue triangles), the resistance attains its maximum for $B$ aligned with the $c$ axis ($\alpha=\ang{0}$) and decreases markedly as $B$ is tilted toward the $ab$ planes. reaching a minimum at $\alpha = \ang{90}$. This behavior is characteristic of anisotropic superconductors, where intrinsic vortex pinning by the layered structure becomes effective for $B \parallel ab$, also known as `vortex lock-in' \cite{TACH89,FEIN90}. The curve qualitatively resembles results on unpatterned Bi-2212 thin films \cite{RAFF91a}, with one exception: The two dips marked with asterisks at $\alpha \sim \ang{79}$ and $\alpha \sim \ang{101}$ are caused by commensurability effects of the perpendicular component of the magnetic field, $B_\parallel$, and will be discussed below.

However, when the first matching field $B_\mathrm{m} = \qty{0.21}{\T}$ is applied and rotated, the red circles in Fig.~\ref{fig:tilt} reveal drastically different behavior. The $R(\alpha)$ curve shows a relatively flat region for $\alpha \leq \ang{10}$, suggesting robust pinning and vortex stiffness. Here, the pancake vortices remain aligned with the columnar defects. For larger values of $\alpha$, the orthogonal field component $B \parallel c$ no longer satisfies the matching condition. Consequently, the stack of pancake vortices undergoes partial dissociation, which enhances vortex mobility and thus increases the electrical resistance, up to a maximum in $R(\alpha)$ at $\alpha \approx \ang{50}$. As $\alpha$ approaches \ang{90}, the resistance drops sharply due to dominating intrinsic pinning \cite{TACH89}. Notably, at $B_\mathrm{m}$, the magnetoresistance ratio $R(\ang{0})/R(\ang{90}) \sim 2.3$ is small, indicating that the pinning of vortices at the CDs has a strength comparable to that of intrinsic pinning by layers. In contrast, at the higher field $B = \qty{1.0}{\T}$, the ratio is $\sim 59$, indicating the presence of mobile interstitial vortices at $\alpha =\ang{0}$.

\begin{figure*}
    \centering
    \includegraphics[width=0.8\linewidth]{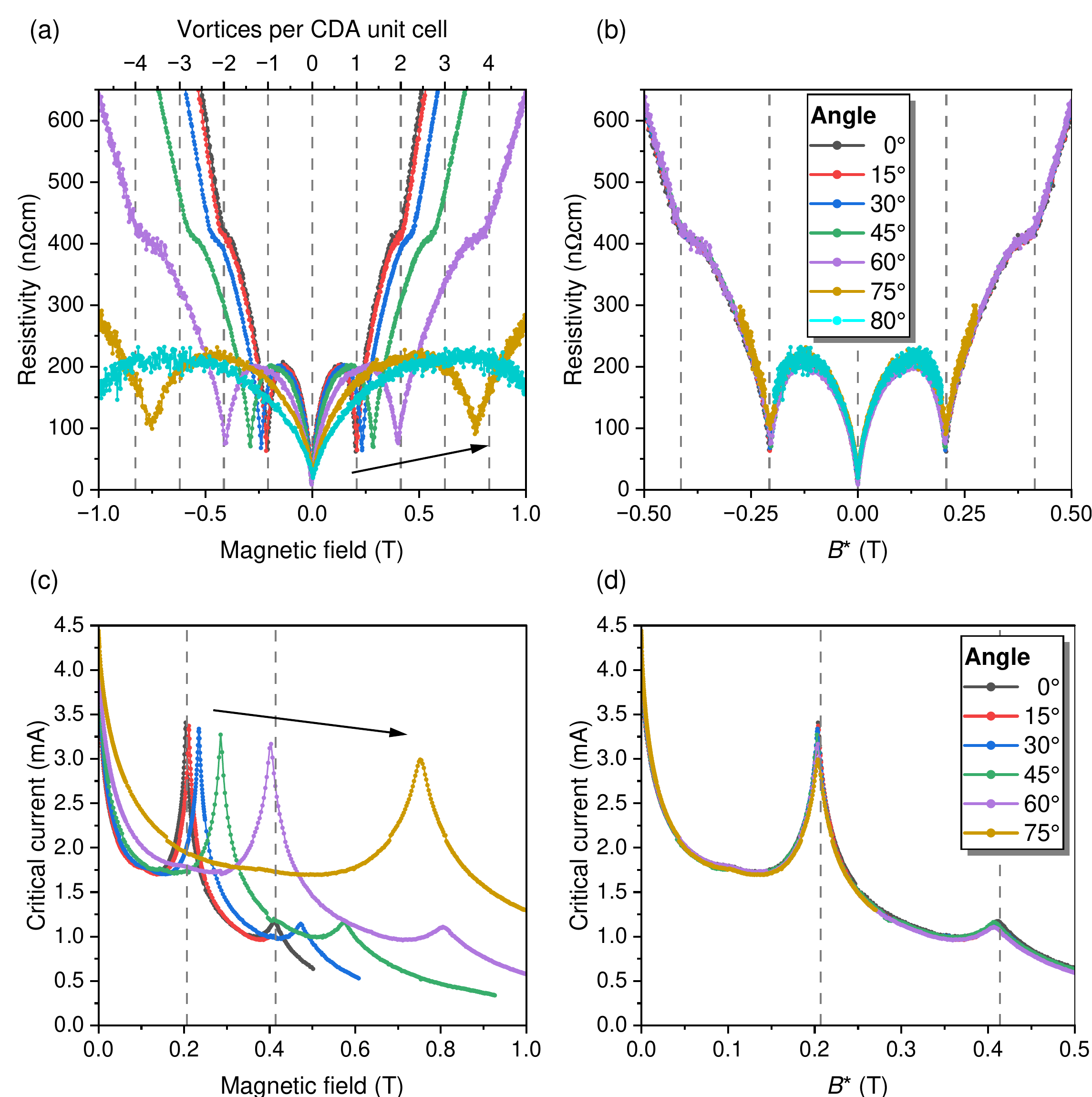}
    \caption{Magnetoresistivity at \qty{33}{\K} and critical current vs magnetic field at \qty{20}{\K} of the Bi-2212 microbridge with a square lattice of \qty{40}{ki/spot} irradiated CDs. Broken lines indicate integer matching fields at $B \parallel c$. (a) $\rho(B)$ curves at magnetic fields tilted by different angles $\alpha$ from \ang{0} to \ang{80}. The arrow indicates the only minor increase of resistivity at the matching dips with tilt angle. (b) The same data plotted against the effective magnetic field $B^*$ of Eq.~\eqref{eq:anisotropy}, showing perfect data collapse onto a universal curve. The legend applies also to panel (a).  (c) $I_c(B)$ curves, for field angles $\alpha$ from \ang{0} to \ang{75}. The arrow indicates the small decrease of critical current maxima with $\alpha$. (d) The data from panel (c) plotted as a function of $B^*$. The legend applies also to panel (c).}
    \label{fig:B_alpha}
\end{figure*}

For further analysis of the anisotropic transport behavior of the \qty{40}{ki/spot}-irradiated Bi-2212 bridge, Fig.~\ref{fig:B_alpha}(a) presents a series of magnetoresistance measurements as sweeps of the applied magnetic field at different fixed angles $\alpha$, again at $T = \qty{33}{\K}$. The positions of the matching minima shift to higher applied magnetic fields, as indicated by the arrow. Remarkably, the resistance value at the first matching minimum remains almost constant over a wide angular range $\alpha = \ang{0} \dots \ang{75}$, indicating that the pinning by the columnar defects remains efficient even when the field is strongly tilted.

The anisotropic superconductor model \cite{BLAT92} predicts that properties should scale by replacing the magnetic field $B$ with an effective field
\begin{equation}
B^*=B \sqrt{\gamma^{-2} \sin ^2 \alpha+\cos ^2 \alpha}\;,
\label{eq:anisotropy}
\end{equation}
where $\gamma$ is an effective anisotropy. However, this model was devised for unpatterned anisotropic superconductors and primarily for magnetic fields near $\alpha = \ang{90}$, where the \emph{intrinsic} pinning dominates.

A key finding is presented in Fig.~\ref{fig:B_alpha}(b), where the data from panel (a) are plotted against the effective magnetic field $B^*$, using $\gamma = 12$ as a fit parameter. The entire set of magnetoresistivity curves taken at different angles collapses perfectly onto a single master curve and the matching minima appear at the same $B^*$. This scaling behavior indicates that the CDA introduces an additional \emph{extrinsic} anisotropy into the system, apart from the intrinsic one defined by the layered structure of Bi-2212. Since $\gamma$ enters Eq.~\eqref{eq:anisotropy} squared, it implies that the vortex dynamics and pinning are predominantly determined by the field component aligned with the columnar defects, $B_\parallel$. Even in oblique magnetic fields, the majority of pancake vortices assemble themselves commensurately along the CDs. Surprisingly, this scaling contrasts with observations in the far less anisotropic compound YBCO, where only the positions of the matching dips scale, while the overall shapes of the $\rho(B)$ curves do not collapse as perfectly \cite{AICH23}.

\subsection{Critical Current Density}

\begin{figure*}
    \centering
    \includegraphics[width=0.8\linewidth]{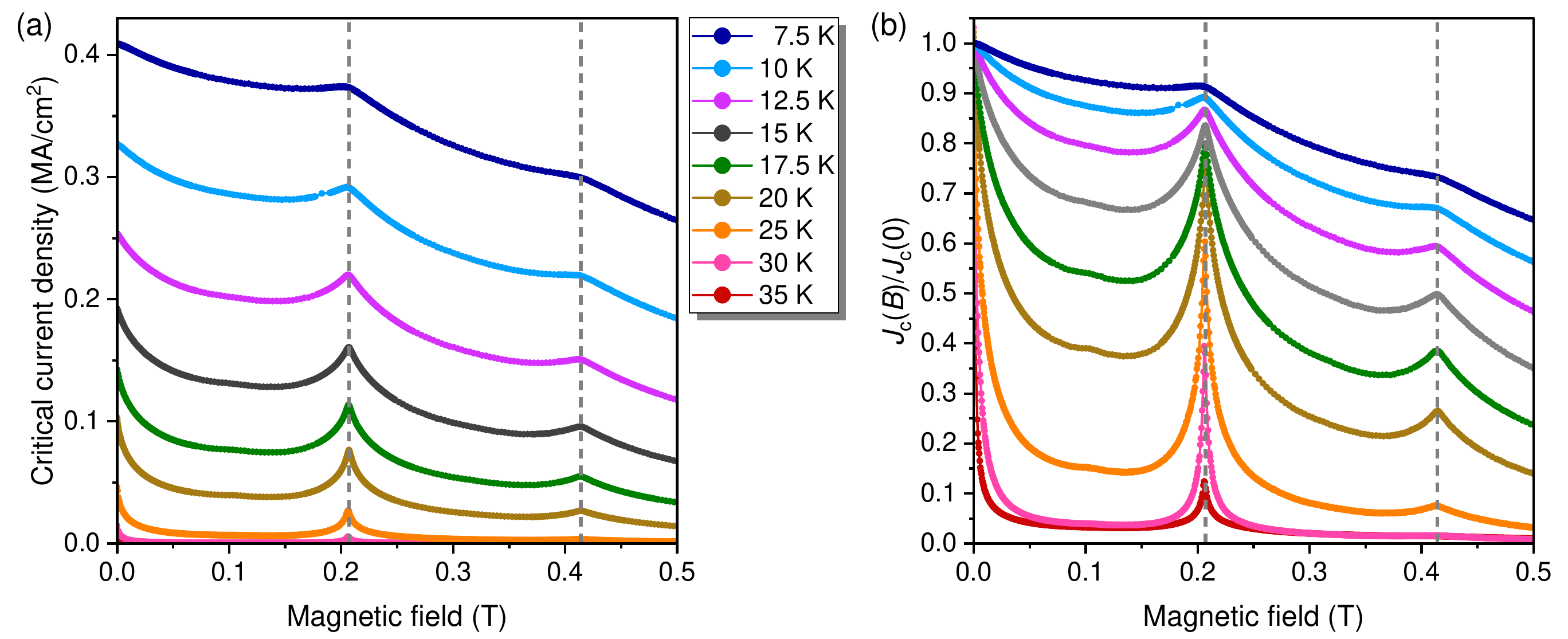}
    \caption{(a) Critical current density $J_\mathrm{c}$ as a function of magnetic field, applied parallel to the $c$ axis of a Bi-2212 microbridge, irradiated with a \qty{40}{ki/spot} square lattice of CDs. (b) Normalized critical current density $J_\mathrm{c}(B)/J_\mathrm{c}(0)$. Color coding is the same for both panels and broken lines indicate the first and second matching fields.}
    \label{fig:Jc}
\end{figure*}

\begin{figure}
    \centering
    \includegraphics[width=0.7\linewidth]{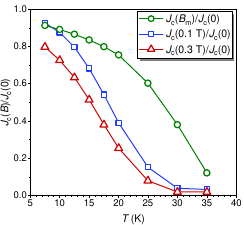}
    \caption{Normalized critical current densities of the bridge irradiated with \qty{40}{ki/spot} at three different magnetic fields: at the matching field $B_\mathrm{m}$ (circles) and at off-matching fields below $B_\mathrm{m}$ (\qty{0.1}{\T}, squares) and above (\qty{0.3}{\T}, triangles).}
    \label{fig:peaks}
\end{figure}

The critical current density $J_c(B)$ of the nanopatterned Bi-2212 epitaxial film with \qty{40}{ki/spot} irradiated CDs also exhibits clear matching effects. Fig.~\ref{fig:Jc}(a) shows $J_c(B \parallel c)$ at various temperatures. Clear peaks in $J_c(B)$ are observed at the first matching field $B_\mathrm{m}$ and moderate ones or shoulders at the second matching field $2 B_\mathrm{m}$. The evolution of the normalized critical current density, $J_c(B)/J_c(B=0)$, displayed in Fig.~\ref{fig:Jc}(b), highlights the prominence of the matching peaks. Particularly in the temperature range of \qty{15}{\K} to \qty{30}{\K}, the peaks are sharp and well-defined, confirming strong commensurability effects. It is informative to compare the relative drop of $J_c(B)/J_c(0)$ for two different fields. In Fig.~\ref{fig:peaks}, normalized $J_c(B_\mathrm{m})/J_c(B=0)$ values are plotted at the matching field $B_\mathrm{m}$ and compared to a smaller ($\qty{0.1}{\T}$) and a larger off-matching field ($\qty{0.3}{\T}$). The graph illustrates the pronounced enhancement of pinning strength arising from the commensurability between the vortices and the CD array.

The angular dependence of the critical current $I_c(\alpha)$ measured at \qty{20}{\K}, shown in Fig.~\ref{fig:tilt}(b), provides strong evidence consistent with the conclusions drawn from magnetoresistance data $R(\alpha)$. At the matching field, $I_\mathrm{c}(\alpha)$ exhibits a plateau in the vicinity of $\alpha = \ang{0}$, which is indicative of enhanced vortex-line stiffness, and attains a magnitude comparable to that of the pronounced intrinsic peak at $\alpha = \ang{90}$. At the larger field of $B = \qty{1.0}{\T}$, the asterisk-labelled maxima arise from commensurability effects associated with the orthogonal field component $B_\parallel$. The weak feature marked by a plus sign is attributed to matching corresponding to two flux quanta per CDA unit cell. Notably, these signatures are more pronounced than the magnetoresistance minima observed in Fig.~\ref{fig:tilt}(a).

Finally, Fig.~\ref{fig:B_alpha}(c) presents the critical current measured at $T = \qty{20}{\K}$ as a function of magnetic field for several tilt angles $\alpha$. Analogous to the behavior observed in the resistivity minima [Fig.~\ref{fig:B_alpha}(a)], the critical-current maxima shift toward higher magnetic fields as the field is rotated away from the $B \parallel c$ configuration. With increasing inclination, the maximum critical current is only weakly reduced, as indicated by the arrow. Upon rescaling the applied field to the effective magnetic field $B^*$, all datasets collapse onto a single curve [Fig.~\ref{fig:B_alpha}(d)] using the same anisotropy parameter $\gamma = 12$. These results independently corroborate the magnetoresistivity measurements and the conclusions drawn from them.

\section{Conclusion}

This study revealed He-FIB as a viable technique for nanopatterning Bi-2212 thin films, opening avenues for studying fundamental vortex physics in this highly anisotropic system with controlled disorder. High-quality, $c$-axis-oriented Bi-2212 thin films were fabricated by pulsed laser deposition followed by an optimized post-annealing process. Focused ion beam irradiation employing \qty{30}{\keV} \ce{He^+} ions created square arrays of nanoscale defect columns with \qty{100}{\nm} period, using fluences ranging from \qty{5}{ki/spot} to \qty{40}{ki/spot}. The efficacy of this process in locally suppressing the critical temperature was demonstrated by magnetotransport measurements revealing clear vortex matching effects, characterized by minima in magnetoresistance and pronounced peaks in the critical current at commensurate magnetic fields. The number of ions per spot required to create comparable matching effects and pinning strengths in the Bi-2212 films was about two to four times larger than for YBCO films (field orientation $B \parallel c$).

The angular dependences of the magnetoresistance and critical current displayed robust scaling within a modified anisotropic superconductor framework when the data were recast as a function of an effective magnetic field, $B^*$. An effective anisotropy parameter, $\gamma = 12$, which combines layer and CD anisotropies, provided a consistent description of both datasets, implying that the vortex dynamics are primarily controlled by the magnetic-field component parallel to the columnar defects. Notably, even at large inclination angles, the maxima of the critical current decreased only marginally. This behavior is consistent with strong confinement of pancake vortices along the irradiation-induced tracks, despite the pronounced intrinsic anisotropy of the material and the expected high mobility of pancake vortices.

These findings underscore the distinct characteristics of vortex matter in highly anisotropic Bi-2212 under an artificial periodic pinning potential. While He-FIB provides a powerful means to control vortex behavior, the consequences of this study reach further, as it shows the path for nanoscale patterning of Bi-2212 thin films---an important prerequisite for devices \cite{DOBR26R} based on this extremely anisotropic and delicate material.

\section{Methods}
\subsection{Film Fabrication and Characterization}

Thin films of \ce{Bi2Sr2CaCu2Ox} were grown on (001)-oriented \ce{LaAlO3} (LAO) substrates by pulsed laser deposition (PLD) from stoichiometric Bi-2212 targets, structured into Hall bars and post-annealed in an oxygen atmosphere. The thickness of the films used in this study was approximately \qty{42}{\nm} (chip A) and \qty{49}{\nm} (chip B).
% chip A = SK060723a; chip B = SK230823a

Details on fabrication, characterization, and electrical magneto-transport properties of similar unpatterned films have been published elsewhere \cite{KEPP23,AICH25}. Briefly, X-ray diffraction (XRD) confirmed the predominant Bi-2212 phase with strong $c$ axis orientation perpendicular to the substrate surface. Moreover, transmission electron microscopy (TEM) demonstrated the layered structure, characteristic of Bi-2212, and a sharp interface with the substrate, with no evidence of significant impurity phases. 

Electrical contact pads were formed by evaporating Ag/Au pads through a copper shadow mask. Subsequently, the Bi-2212 films were patterned into four rectangular microbridges with side arms on each chip using UV photolithography and wet chemical etching and annealed once more in oxygen atmosphere. The dimensions are \qty{102}{\um} (\qty{98}{\um}) width and \qty{497}{\um} (\qty{501}{\um}) distance between the voltage probes for bridges on chip A (chip B). Electrical contacts in the cryostat were established using silver paste and $50\,\mu$m thick gold wires.

\subsection{Focused Helium Ion Beam (He-FIB) Irradiation}

To create a periodic array of artificial pinning centers, the Bi-2212 thin films were irradiated with a \qty{30}{\keV} focused \ce{He^+} ion beam in a helium-ion microscope (Zeiss Orion NanoFab). The instrument was set to \qty{175}{\um} field-of-view, \qty{20}{\um} aperture, spot control 4.5 and \qty{5.6}{\pA} (\qty{3.2}{\pA}) beam current for chip A (chip B), delivering a beam with an estimated FWHM of about \qty{21}{\nm}. These settings ensured the creation of highly localized defects at each irradiated spot. A square lattice of defect sites with a period of $a = \qty{100}{\nm}$ was written into each bridge by irradiating three nonoverlapping areas of \qty{120}{\um} width and \qty{150}{\um} length. The number of ions per spot was specific for every bridge, \qty{5}{ki/spot} and \qty{10}{ki/spot} on chip A, \qty{20}{ki/spot} and \qty{40}{ki/spot} on chip B. 
\\

\subsection{Magnetotransport Measurements}

The bridges' magnetoresistance $R(T,B)$ as a function of temperature $T$ and magnetic field $B$ was measured in standard four-probe geometry in a pulse-tube cryostat using an external constant-current source (Keithley 6221) and a nanovoltmeter (Keithley 2182) and collecting data at both current polarities to exclude thermovoltages. The magnetic field, provided by an electromagnet, was rotated by an angle $\alpha$ in a plane perpendicular to the injected current, where $\alpha = \ang{0}$ corresponds to the magnetic field vector parallel to the crystallographic $c$ axis.

The critical current $I_c$ was determined under isothermal, constant-magnetic-field conditions from current-voltage characteristics by increasing the current in small steps up to an electric-field criterion of $10\,\mu$V/cm and interpolating between data points. 

\subsection*{Data availability}
The data generated and analyzed during this study are available from the corresponding author upon reasonable request.

\bibliography{nano2212, references}

\subsection*{Acknowledgements}
This research was funded in whole, or in part, by the Austrian Science Fund (FWF) Grant-DOI: 10.55776/I4865 and by the German Research Foundation (DFG), grants KO~1303/16-1 and Go~1106/7-1. For the purpose of open access, the authors have applied a CC BY public copyright license to any Author Accepted Manuscript version arising from this submission. The publication is based upon work from COST Actions CA21144 (SuperQuMap) and CA23134 (Polytopo) supported by COST (European Cooperation in Science and Technology).

\subsection*{Author contributions}
W.L., B.A., J.D.P., and D.K. conceived the experiments; S.K. and J.D.P. fabricated and characterized the samples; C.S. and E.G. irradiated the samples; B.A. performed the transport measurements; B.A. and W.L. analyzed the results. W.L. drafted the manuscript, and all authors contributed to the manuscript and reviewed it.

\end{document}